\documentclass[
twocolumn,
reprint,
superscriptaddress,
amsmath,
amssymb,
aps,
pra,
]{revtex4-1}
 
\usepackage[latin1]{inputenc}
\usepackage{amsmath}
\usepackage{amsfonts}
\usepackage{hyperref}
\usepackage{amssymb}
\usepackage{graphicx}
\usepackage{hyperref}
\usepackage{colortbl}
\usepackage{subfigure}
\usepackage{svg}

\usepackage[normalem]{ulem}

\usepackage[T1]{fontenc}
\usepackage{mathptmx}
\usepackage{bbold}
\usepackage{scalerel}
\usepackage{xcolor}

\usepackage{mathtools}

\newcommand*\ch[1]{\ensuremath{\mathrm{#1}}}
\newcommand{\hyref}[1]{\hyperref[#1]{\ref{#1}}}
\newcommand{\dd}{\mathrm{d}}

\newcommand{\orange}[1]

\begin{document}

\title{Unveiling the capabilities of bipolar conical channels in neuromorphic iontronics}
\author{T. M. Kamsma}
\affiliation{Institute for Theoretical Physics, Utrecht University,  Princetonplein 5, 3584 CC Utrecht, The Netherlands}
\affiliation{Mathematical Institute, Utrecht University, Budapestlaan 6, 3584 CD Utrecht, The Netherlands}
\author{W. Q. Boon}
\affiliation{Institute for Theoretical Physics, Utrecht University,  Princetonplein 5, 3584 CC Utrecht, The Netherlands}
\author{C. Spitoni}
\affiliation{Mathematical Institute, Utrecht University, Budapestlaan 6, 3584 CD Utrecht, The Netherlands}
\author{R. van Roij}
\affiliation{Institute for Theoretical Physics, Utrecht University,  Princetonplein 5, 3584 CC Utrecht, The Netherlands}

\date{\today}

\begin{abstract}
Conical channels filled with an aqueous electrolyte have been proposed as promising candidates for iontronic neuromorphic circuits. This is facilitated by a novel analytical model for the internal channel dynamics [Kamsma et al., arXiv:2301.06158, 2023], the relative ease of fabrication of conical channels, and the wide range of achievable memory retention times by varying the channel lengths. In this work, we demonstrate that the analytical model for conical channels can be generalized to channels with an inhomogeneous surface charge distribution, which we predict to exhibit significantly stronger current rectification and more pronounced memristive properties in the case of bipolar channels, i.e.\ channels where the tip and base carry a surface charge of opposite sign. Additionally, we show that the use of bipolar conical channels  in a previously proposed iontronic circuit features hallmarks of neuronal communication, such as all-or-none action potentials and spike train generation. Bipolar channels allow, however, for circuit parameters in the range of their biological analogues, and exhibit membrane potentials that match well with biological mammalian action potentials, further supporting its potential for bio-compatibility.
\end{abstract}

\maketitle
\section{Introduction}
Iontronics is an exciting emerging platform that aims to harness the transport of ions for the use of signalling. The ability to control ion transport in confined geometries offers unique opportunities compared to traditional electronic systems, such as the ability to mimic biological processes or interface with cells and tissues \cite{han2022iontronics,yang2019cavity,hu2019ultrasensitive}. A particularly exciting direction is that of neuromorphic (brain-inspired) iontronic circuits \cite{han2022iontronics,yang2019cavity,hu2019ultrasensitive,noy2023nanofluidic,li2020synaptic,xie2022perspective}, which offer the unique feature of closely resembling biological processes as signalling in the brain also relies on ion transport \cite{fundNeuroE,sah2014brains}. A promising candidate for the realisation of such circuits are ionic microfluidic memristors (memory-resistors) \cite{wang2012transmembrane, li2015history, wang2014physical, wang2016dynamics, wang2017correlation, sheng2017transporting, brown2020deconvolution, brown2022selective, brown2021higher, wang2018hysteresis,  ramirez2021negative,sun2015fluidic,robin2021principles,robin2023long,xiong2023neuromorphic,kamsma2023iontronic}. The dynamical properties of memristors make them artificial analogues to biological synapses, the connections between neurons \cite{chua2013memristor,van2018organic,keene2021neuromorphic, chicca2020recipe,christensen20222022}. Over the past few years, a vast array of different memristors has been extensively investigated as components for neuromorphic circuit architectures \cite{schuman2017survey,venkatesan2022brain,zhu2020comprehensive}. Not only are memristors analogues to synapses, the biological ion channels present in neuronal membranes that facilitate signalling \cite{lucas1909all,bean2007action,fundNeuroE,fundNeuroTrain,cymbalyuk2002bursting,marder2002cellular,sherman2001tonic,hodgkin1952quantitative,rall2011core,fitzhugh1973dimensional,rall1962theory,halter1991distributed,hay2011models,hines1997neuron,kole2008action} are also memristors \cite{sah2014brains,chua2013memristor}, offering even more perspectives for brain-inspired circuits. Due to the prospect of more energy-efficient computers \cite{mehonic2022brain,sangwan2020neuromorphic} and bio-compatibility \cite{keene2020biohybrid,harikesh2022organic,krauhausen2021organic,marasco2021neurorobotic,van2018organic,yuan2021organic}, memristors and neuromorphic circuit architectures have drastically increased in popularity over recent years \cite{schuman2017survey,venkatesan2022brain,zhu2020comprehensive}. However, the emphasis has mostly been on memristors that require electrons or holes as charge carriers \cite{schuman2017survey,venkatesan2022brain,sangwan2020neuromorphic,zhu2020comprehensive}, limiting their applicability in fully ionic fluidic systems.

In the past few years, however, some interest has been shown in microfluidic neuromorphic components \cite{kamsma2023iontronic,robin2021principles,robin2023long,xiong2023neuromorphic,wang2012transmembrane, li2015history, wang2014physical,  wang2016dynamics, wang2017correlation, sheng2017transporting, brown2020deconvolution, brown2022selective, brown2021higher, wang2018hysteresis, ramirez2021negative}. Candidates of specific interest for the present work are conical channels containing an aqueous electrolyte and a homogeneous surface charge, which are known to act as iontronic microfluidic memristors \cite{wang2012transmembrane, li2015history, wang2014physical,  wang2016dynamics, wang2017correlation, sheng2017transporting, brown2020deconvolution, brown2022selective, brown2021higher, wang2018hysteresis, ramirez2021negative}. Recently, an analytical model was derived that explains in a quantitative manner how transient concentration polarisation in such channels is responsible for a volatile conductance memory and it was demonstrated that these channels could carry the potential to be used in experimentally accessible neuromorphic iontronic circuits \cite{kamsma2023iontronic}. The underlying functionality which underpins the memristive behaviour of conical channels is that they exhibit current rectification in steady-state \cite{wei1997current, boon2021nonlinear, white2008ion, jubin2018dramatic, vlassiouk2009biosensing}. Although conical channels with a homogeneous surface charge distribution are desirable as relatively simple model systems for investigating iontronic systems \cite{cheng2007rectified, siwy2006ion, bush2020chemical, jubin2018dramatic, siwy2002rectification,fulinski2005transport,siwy2005asymmetric, duleba2022effect, lan2016voltage, vlassiouk2008nanofluidic, liu2012surface, kubeil2011role, boon2021nonlinear, dal2019confinement, poggioli2019beyond,uematsu2022analytic}, they are not necessarily the best performing channels for current rectification \cite{huang2018bioinspired}. In fact, conical channels with a surface charge distribution that changes sign as a function of the distance to the tip are known to exhibit a much stronger current rectification than homogeneously charged ones \cite{vlassiouk2007nanofluidic}. These so-called bipolar conical channels are therefore promising systems to advance the field of iontronic (neuromorphic) systems.

In this work we present an analytical model that quantitatively captures both the steady-state and the dynamical behaviour of conical channels with an inhomogeneous surface charge distribution, based on the methodology in Refs.~\cite{boon2021nonlinear,kamsma2023iontronic}. Our model contains no free parameters and can quantitatively predict the steady-state and time-dependent ionic charge currents as a result of static and dynamic applied voltages, respectively. An understanding of these current-voltage relations and the dependence on system parameters could allow for a targeted development of new circuits of these channels and more effortless scanning of possible applications thereof.

Recently some fully microfluidic circuits that display neuronal behaviour have been theoretically and numerically explored. These circuits, through which an imposed current can be driven, consist of artificial ion channels, batteries, and a capacitor \cite{robin2021principles,kamsma2023iontronic}. In Ref.~\cite{robin2021principles} a circuit was modelled containing quasi two-dimensional nanochannels that connect aqueous electrolytes, which generated a train of voltage spikes, a feature of neuronal communication \cite{fundNeuroTrain,cymbalyuk2002bursting,marder2002cellular,sherman2001tonic,bean2007action}. In Ref.~\cite{kamsma2023iontronic}, a more experimentally accessible circuit containing conical channels with homogeneous surface charge was proposed that also obeys the defining all-or-none law for action potentials \cite{lucas1909all,bean2007action,fundNeuroE}. Here we will demonstrate that bipolar conical channels can be used in the circuit from Ref.~\cite{kamsma2023iontronic} to achieve the same all-or-none action potentials and spike trains, however with more biologically relevant salt concentrations and battery potentials. Furthermore, the circuit's membrane potentials during spiking match typical mammalian values, making it more bio-compatible.

\section{Bipolar conical channel}
\begin{figure}[ht]
	\centering
	\includegraphics[width=0.42\textwidth]{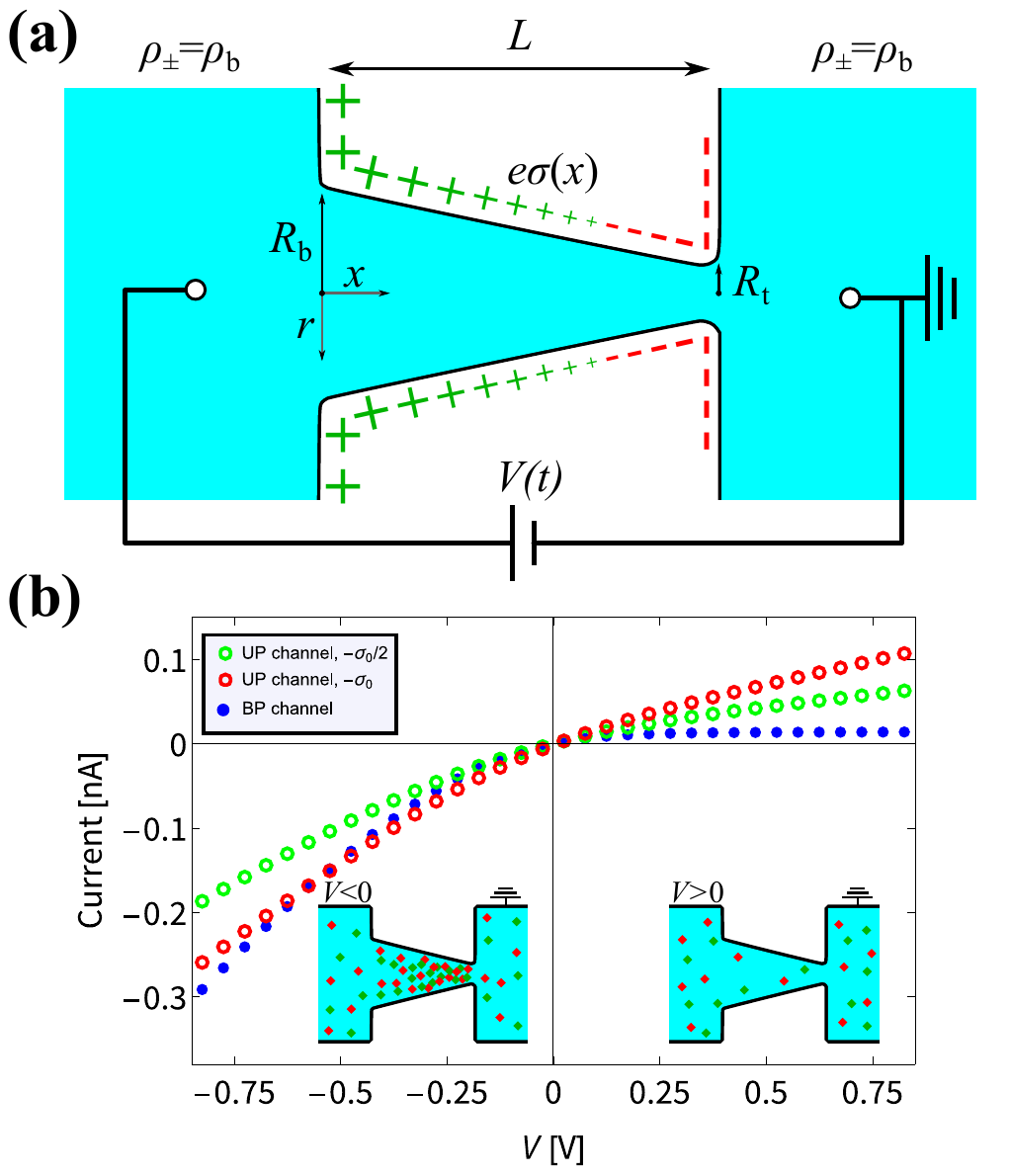}\vspace{-0.4 cm}
	\caption{\textbf{(a)} Schematic representation of the azimuthally symmetric bipolar (BP) conical channel (not to scale), with channel length $L$, base radius $R_{\mathrm{b}}$, and tip radius $R_{\mathrm{t}}<R_{\mathrm{b}}$, connecting two bulk reservoirs of a 1:1 aqueous electrolyte, with bulk concentration $\rho_{\mathrm{b}}$. The channel wall carries a surface charge density $e\sigma(x)$, with $\sigma(x)=\sigma_{0}+\sigma^{\prime}\frac{x}{L}$. Here $\sigma^{\prime}=-3\sigma_0/2$ with $e\sigma_0=0.1\;e\text{nm}^{-2}$ such that the surface charge is positive at the base ($e\sigma(0)=e\sigma_0$) and negative at the tip $(e\sigma(L)=-e\sigma_0/2)$. A possibly time-dependent electric potential drop $V(t)$ is applied over the channel, driving an ionic charge current $I(t)=g(V(t),t)V(t)$ with $g(V(t),t)$ the channel conductance that we calculate in this paper. \textbf{(b)} Steady-state current $I$ as a function of the static potential $V$ as predicted by full FE calculations of the PNPS equations (\ref{eq:ce})-(\ref{eq:poisson}), for a bipolar (BP) channel (blue) and otherwise identical unipolar (UP) channels with uniform surfaces charge $-\sigma_0/2$ (green) and $-\sigma_0$ (red). An applied positive (negative) voltage over the channel results in ion depletion (accumulation) as depicted in the insets of \textbf{(b)}, responsible for the steady-state diodic behaviour of the cones \cite{boon2021nonlinear}.}
	\label{fig:setup}
\end{figure}

To study the steady-state properties and the voltage-driven dynamics of conical channels with an inhomogeneous surface charge $e\sigma(x)$, we consider an azimuthally symmetric long conical channel of length $L$, base radius $R_{\ch{b}}\ll L$ at $x=0$, and tip radius $R_{\ch{t}}<R_{\ch{b}}$ at $x=L$, with $x$ the longitudinal coordinate, as depicted schematically in Fig.~\ref{fig:setup}(a). 
The channel radius is described by $R(x)=R_{\mathrm{b}}-x\Delta R/L$ for $x\in\left[0,L\right]$, the central axis being at radial coordinate $r=0$ and $\Delta R=R_{\mathrm{b}}-R_{\mathrm{t}}$. On the channel surface, at $r=R(x)$, we assume an inhomogeneous surface charge distribution $e\sigma(x)$, with a positive surface charge at the base and middle of the channel, a negative surface charge at the tip, with a linear increase described by 
\begin{equation}\label{eq:sigma}
    \sigma(x)=\sigma_{0}+\sigma^{\prime}\frac{x}{L},
\end{equation}
where throughout this manuscript we set $\sigma^{\prime}=-3\sigma_0/2$ with $e\sigma_0=0.1\;e\text{nm}^{-2}$, resulting in a bipolar (BP) channel. Unless stated otherwise, we set the channel dimensions as length $L=10\text{ }\mu\text{m}$, base radius $R_{\mathrm{b}}=200\text{ nm}$, and tip radius $R_{\mathrm{t}}=50\text{ nm}$, resulting in a channel geometry similar as realised before experimentally \cite{kovarik2009effect}. The channel connects two bulk reservoirs of an incompressible aqueous 1:1 electrolyte, with mass density $\rho_{\mathrm{m}}=10^3\text{ kg}\cdot\text{m}^{-3}$, viscosity $\eta=1.01\text{ mPa}\cdot\text{s}$, and electric permittivity $\epsilon=0.71\text{ nF}\cdot\text{m}^{-1}$. The electrolyte contains ions with monovalent charges $\pm e$ with $e$ the proton charge and diffusion coefficients $D_{\pm}=D=2\text{ }\mu\text{m}^2\text{ms}^{-1}$, a typical value for dilute KCl \cite{lide2004crcKCl} in microfluidic channels \cite{choi2016high,shen2010microfluidic}. At the far side of both reservoirs we impose fixed ion concentrations $\rho_{\pm}=\rho_{\mathrm{b}}=2\text{ mM}$, such that the equilibrium Gouy-Chapman surface potential $\psi_0(x)=(2k_{\ch{B}}T/e)\sinh^{-1}[2\pi\lambda_{\ch{B}}\lambda_{\ch{D}}\sigma(x)]$ equals $\psi_0(0)\approx 92\text{ mV}$ at the base and $\psi_0(L)\approx-61\text{ mV}$ at the tip. Here we introduced the Bjerrum length $\lambda_{\ch{B}}=e^2/(4\pi\epsilon k_{\ch{B}}T)$  and the Debye length $\lambda_{\ch{D}}=1/\sqrt{8\pi\lambda_B\rho_{{\ch{b}}}}$. An electric double layer (EDL) forms that screens the surface charge with  $\lambda_{\mathrm{D}}\approx6.8\text{ nm}$. The far sides of both reservoirs are kept at a constant and equal pressure $P=P_0$. On the far side of the reservoir connected to the base we impose an electric potential $V(t)$, while the far side of the other reservoir is grounded. The resulting physical quantities of interest in this system are the electric potential profile $\Psi(x,r,t)$, the ionic concentration profiles $\rho_{\pm}(x,r,t)$, an electro-osmotic fluid flow with velocity field $\mathbf{u}(x,r,t)$, ionic fluxes $\mathbf{j}_{\pm}(x,r,t)$ with $\mathbf{j}_{+}-\mathbf{j}_{-}$ the charge flux and $\mathbf{j}_{+}+\mathbf{j}_{-}$ the salt flux, and the pressure profile $P(x,r,t)$.

The aforementioned physical quantities can be described by a coupled set of equations. Firstly, the ionic fluxes $\mathbf{j}_{\pm}$ and concentration profiles $\rho_{\pm}$ satisfy the continuity equation
\begin{gather}
	\dfrac{\partial\rho_{\pm}}{\partial t}+\nabla\cdot\mathbf{j}_{\pm}=0,\label{eq:ce}
\end{gather}

and the Nernst-Planck equation,
 \begin{gather}
 \mathbf{j}_{\pm}=-D_{\pm}\left(\nabla\rho_{\pm}\pm\rho_{\pm}\frac{e\nabla \Psi}{k_{\mathrm{B}}T}\right)+\mathbf{u}\rho_{\pm},\label{eq:NP}
\end{gather}
where the three terms account for Fickian diffusion, Ohmic conduction, and Stokesian convection, respectively. The fluid flow $\mathbf{u}(x,r,t)$ satisfies a force balance described by the Stokes equation for an incompressible fluid
\begin{gather}
	\rho_{\mathrm{m}}\dfrac{\partial\mathbf{u}}{\partial t}=\eta\nabla^2\mathbf{u}-\nabla P-e\rho_{\mathrm{e}}\nabla \Psi;\qquad\nabla\cdot\mathbf{u}=0,\label{eq:stokes}
\end{gather}
where $-e\rho_{\mathrm{e}}\nabla\Psi$ is the electric body force, which depends on the ionic space charge density $\rho_{\mathrm{e}}=\rho_{+}-\rho_{-}$. This space charge density affects the electric potential $\Psi(x,r,t)$, that satisfies the Poisson equation
\begin{align}
	\nabla^2\Psi=-\frac{e}{\epsilon}\rho_{\mathrm{e}},\label{eq:poisson}
\end{align}
where $\Psi(-\infty,r,t)=V(t)$ and $\Psi(\infty,r,t)=0$. Eqs.~(\ref{eq:ce})-(\ref{eq:poisson}) form the Poisson-Nernst-Planck-Stokes (PNPS) system of equations. To make the system closed we impose the boundary conditions on the channel wall given by the no-slip condition $\mathbf{u}=0$, the blocking condition $\mathbf{n}\cdot\mathbf{j}_{\pm}=0$, and Gauss' law $\mathbf{n}\cdot\nabla\Psi(x,R(x),t)=e\sigma(x)/\epsilon$, with $\mathbf{n}$ the normal vector of the wall.

In Fig.~\ref{fig:setup}(b) we show steady-state current-voltage (I-V) curves as determined by finite-element (FE) calculations, not only for the BP channel under consideration (blue) but also for homogeneous and unipolar (UP) surface charge densities $-e\sigma_0/2$ (green) and $-e\sigma_0$ (red). An applied potential $V$ over the channel leads to a depletion or accumulation of ions, where for our parameters $V<0$ results in salt accumulation while $V>0$ depletes the channel of salt, as shown in the insets of Fig.~\ref{fig:setup}(b), thereby changing the channel conductance. This concentration polarisation is responsible for the ion current rectification found in conical channels \cite{boon2021nonlinear}. It is clear that the BP channel exhibits a significantly stronger current rectification, with the ratio $ICR=|I(-0.8 \text{ V})/I(0.8 \text{ V})|$ of the current $I(V)$ at voltages $V=\pm0.8\text{ V}$ being as large as $ICR\approx 21$ for the BP channel (blue), while it is as small as $ICR\approx 3$ and 2.4 for the UP channels with surface charges $-\sigma_0/2$ (green) and $-\sigma_0$ (red), respectively.

\section{Analytic approximation for bipolar channels}\label{sec:AA}
The full PNPS equations (\ref{eq:ce})-(\ref{eq:poisson}) cannot be solved analytically for the system of interest here in steady-state. However, with a few reasonable assumptions we can simplify them to obtain some closed-form analytic descriptions \cite{boon2021nonlinear}. Under the assumption that the Debye length is small compared to the channel radius, i.e.\ $\lambda_{\ch{D}}\ll R(x)$, we can make the approximation that for all $r$ at least a few $\lambda_{\ch{D}}$ away from the surface, the salt concentration $\rho_{+}(x,r)+\rho_{-}(x,r)=\rho_{\ch{s}}(x,r)\approx \overline{\rho}_{\mathrm{s}}(x)$ and the electric potential $\Psi(x,r)\approx \overline{\Psi}(x)$ are radially independent. With this assumption, as in Ref.~\cite{boon2021nonlinear},  the slab-averaged electric field $-\partial_x\overline{\Psi}(x)$ and the total salt flux $\mathbf{j}_{\ch{s}}(x,r)=\mathbf{j}_{+}(x,r)+\mathbf{j}_{-}(x,r)$ can be radially integrated to obtain expressions for the cross-sectional averaged electric field
\begin{equation}\label{eq:Efield}
    -\partial_x\overline{\Psi}(x)=\frac{V}{L}\frac{R_{\ch{b}}R_{\ch{t}}}{R^2(x)},
\end{equation}
and the total salt flux $J_{x}(x)=2\pi\int_0^{R(x)}\dd r\, r\mathbf{j}_{\ch{s}}(x,r)\cdot\hat{\mathbf{x}}$ through the channel
\begin{equation}\label{eq:saltflux}
\begin{split}
    J_x(x)=&-D\left(\pi R^2(x)\partial_x\overline{\rho}_{\mathrm{s}}(x)+2\pi \sigma(x)\frac{eV}{k_{\mathrm{B}}T}\frac{R_{\mathrm{t}}R_{\mathrm{b}}}{R(x)L}\right)\\
    &+Q(V)\overline{\rho}_{\mathrm{s}}(x),
    \end{split}
\end{equation}
with $Q(V)=-\pi R_{\mathrm{t}}R_{\mathrm{b}}\epsilon\psi_{\ch{eff}} V/(\eta L)$ the electro-osmotic fluid volume flow, which is similar to the expression for the fluid flow of a UP channel \cite{boon2021nonlinear} except for the surface potential term $\psi_{\ch{eff}}$. In BP channels it is not immediately clear how the inhomogeneous surface charge dictates the electro-osmotic flow. For our standard parameter set we use $\psi_{\ch{eff}}\approx -25\text{ mV}$ as a fit parameter, which will be discussed further in Sec.\ \ref{sec:FE_ver}. Eq.~(\ref{eq:saltflux}) represents the diffusive, conductive and convective components of the salt flux, respectively. In steady-state the condition $\partial_xJ(x)=0$ must hold, yielding for given $\sigma(x)$, $R(x)$, and $Q(V)$ a differential equation for the unknown radially averaged salt concentration profile function $\overline{\rho}_{\mathrm{s}}(x)$. In Ref.~\cite{boon2021nonlinear} this differential equation is solved for a conical channel with a homogeneous surface charge and boundary conditions $\overline{\rho}_{\mathrm{s}}(0)=\overline{\rho}_{\mathrm{s}}(L)=2\rho_b$. Here we consider the case where the surface charge distribution is given by Eq.~(\ref{eq:sigma}). By solving $\partial_xJ(x)=0$ for a given static potential $V$ we obtain the following expression for the radially averaged salt concentration
\begin{equation}\label{eq:rhos}
\begin{split}
    \overline{\rho}_s(x,V)=2\rho_b-\frac{1}{\text{Pe}/V}\frac{2e \left(\sigma_{0}\Delta R+\sigma^{\prime} R_{\mathrm{b}}\right)}{k_{\mathrm{B}}T  R_{\mathrm{t}}^2}\\
    \left(\frac{R_{\mathrm{b}}(1-x/L)}{R(x)}-\frac{e^{-\text{Pe}\frac{(1-x/L)R_{\mathrm{t}}}{R(x)}}-1}{e^{-\text{Pe}\frac{R_{\mathrm{t}}}{R_{\mathrm{b}}}}-1}\right),
    \end{split}
\end{equation}
with $\text{Pe}=Q(V)L/(\pi DR_{\ch{t}}^2)$ the P\'{e}clet number at the narrow end. Note that for our case with solely a voltage-driven flow without any pressure-driven contribution, $Q(V)=-\pi R_{\mathrm{t}}R_{\mathrm{b}}\epsilon\psi_{\ch{eff}} V/(\eta L)$ is proportional to $V$, and hence the ratio $\text{Pe}/V$ that appears in Eq.~(\ref{eq:rhos}) does not depend on the static potential $V$. In Sec.~\ref{sec:FE_ver} we will see that for our case of $\psi_{\ch{eff}}<0$ a negative applied voltage ($V<0$) will cause an enhancement of the ion concentration in the channel (and hence an increased conductivity), whereas a positive potential ($V>0$) gives rise to an ionic depletion and a reduced conductivity, where the effect of ion accumulation and depletion becomes stronger upon increasing $|V|$. For $V>0$ we will see that the profile as predicted by Eq.~(\ref{eq:rhos}) can even become negative, which is obviously an unphysical result that stems from a break-down of the $\lambda_{\ch{D}}\ll R(x)$ assumption that underlies Eq.~(\ref{eq:rhos}). However, we will discuss in Sec.~\ref{sec:FE_ver} how we can still ensure good agreement on the current-voltage relation over a wide voltage range.

Interestingly, Eq.~(\ref{eq:rhos}) suggests that it can also explain and predict current rectification in cylindrical channels \cite{karnik2007rectification,daiguji2005nanofluidic,meng2015cooperative} as long as $\sigma^{\prime}\neq 0$, since in this case a non-trivial source term remains in Eq.~(\ref{eq:rhos}) even for $\Delta R=0$. Hence our current work suggests to unify the theories for non-linear transport through cylindrical and conical channels carrying homogeneous or inhomogeneous surface charges. Additionally we note that Eq.~(\ref{eq:rhos}) seems to suggest that bipolar and conical rectification mechanisms can oppose each other, even to the extent that no current rectification is expected if $\sigma_0\Delta R=-\sigma^{\prime}R_{\rm{b}}$ (which for our linear surface charge density profile implies $\sigma_{\rm{t}}/\sigma_{\rm{b}}=R_{\rm{t}}/R_{\rm{b}}$ with $\sigma_{\rm{t}}$ and $\sigma_{\rm{b}}$ the surface charge at the tip and the base, respectively). Probing this unification will be left for future work, while we will focus here on a more constrained parameter set to investigate the iontronic neuromorphic circuit in Sec.~\ref{sec:HHcircuit}.

The static electric conductance of the conical channel can be found by treating the concentration profile as a series of resistors of thickness $\dd x$ and cross-sectional area $\pi R^2(x)$. Since the electric field scales with the inverse of $R^2(x)$ according to Eq.~(\ref{eq:Efield}), the contribution to the resistance of each slab equals $\left(g_0\overline{\rho}_{\ch{s}}(x)/(2\rho_{\ch{b}})\right)^{-1}\dd x$ with the homogeneous channel conductance $g_0=(\pi R_{\mathrm{t}} R_{\mathrm{b}}/L)(2\rho_{\rm{b}}e^2D/k_{\mathrm{B}}T)$ \cite{boon2021nonlinear,werkhoven2020coupled}, yielding for the static channel conductivity
\begin{align}\label{eq:cond}
    g_{\infty}(V)=&g_0\frac{L}{2\rho_b\int_{0}^{L}\left(\overline{\rho}_{\ch{s}}(x,V)\right)^{-1}\dd x}.
\end{align}
In order to account for the possibility of unphysical negative concentration profiles at high positive voltages, we replace $\overline{\rho}_{\ch{s}}(x,V)$ by $\max\left[0.2\rho_b,\overline{\rho}_{\ch{s}}(x,V)\right]$ in the actual (numerical) evaluations of Eq.~(\ref{eq:cond}), such that we effectively take the surface conductivity into account by not allowing the concentration profile to drop below 10\% of the bulk salt concentration $2\rho_{\ch{b}}$. This ad hoc cut-off can certainly be improved upon, although the details of the cut-off have limited effects for the system parameters that we use and discuss below. The steady-state current is then given by
\begin{align}\label{eq:Iss}
     I(V)=&g_{\infty}(V)V.
\end{align}
As we will show in Sec.~\ref{sec:FE_ver}, Eq.~(\ref{eq:Iss}) predicts a diodic behaviour of the conical channel through ion depletion (and hence a low conductivity) for $V>0$ and ion accumulation (and hence a high conductivity) for $V<0$.

It was found in Ref.~\cite{kamsma2023iontronic} that the process of ion accumulation and depletion is not instantaneous and occurs over a diffusion-like timescale. To derive an expression for the timescale of this process and thus the typical memory retention time $\tau$ of a BP conical channel from the PNPS equations (\ref{eq:ce})-(\ref{eq:poisson}), we apply the same methodology. We consider two quantities, the total number of ions $N=\pi\int_0^LR^2(x)\overline{\rho}_{\mathrm{s}}(x)\dd x$ and the net salt flux $J_{x}(0)-J_{x}(L)$ into the channel. The change of $N$ given by Eq.~(\ref{eq:rhos}) upon a small voltage perturbation $V^{\prime}$ around $V=0$ yields
\begin{align}\label{eq:alpha}
\left.\dfrac{\partial N}{\partial V}\right|_{V=0} V^{\prime}=\frac{\pi}{6}L\frac{e(\Delta R\sigma_{\ch{0}}+R_{\ch{b}}\sigma^{\prime})}{k_{\ch{B}}T} V^{\prime}\equiv \alpha V^{\prime},
\end{align}
where $\alpha<0$ for the standard parameter set of our BP channel, in agreement with the enhanced (reduced) conductance of a negative (positive) potential $V^{\prime}$. 

At $V=0$ the concentration profile is at equilibrium, so for a small voltage perturbation $V^{\prime}$ we can assume $\bar{\rho}_{\rm{s}}(x)=2\rho_{\rm{b}}$. With this assumption the first and third terms in Eq.~(\ref{eq:saltflux}) vanish. The net salt flux into the channel, $J_x(0)-J_x(L)$, is then determined by the remaining conductive terms
\begin{align}\label{eq:netflux}
    J_x(0)-J_x(L)=2\pi\frac{D}{L}\frac{e(\Delta R\sigma_{0}+R_{\ch{b}}\sigma^{\prime})}{k_{\mathrm{B}}T}V^{\prime}\equiv \gamma V^{\prime},
\end{align}
where $\gamma<0$ for our parameter choices. The typical time it takes for ion depletion or accumulation, and thus the typical memory retention timescale is then approximated by $\tau=\alpha/\gamma$. This yields, perhaps surprisingly, the purely diffusive timescale
\begin{align}
    \tau=\frac{L^2}{12D},\label{eq:ts} 
\end{align}
identical to the expression for UP channels \cite{kamsma2023iontronic}, which is remarkable as the conductive terms in Eq.~(\ref{eq:saltflux}), through which Eq.~(\ref{eq:ts}) is obtained, differ from those in Ref.~ \cite{kamsma2023iontronic}. For our standard parameter set we find $\tau\approx 4.17\text{ ms}$. By assuming that $\partial_tg(V(t),t)\propto g_\infty( V(t))-g(V(t),t)$, an assumption proven to be effective before \cite{markin2014analytical,robin2023long,kamsma2023iontronic}, we can describe the time-dependent conductance $g(V(t),t)$ at a given applied voltage $V(t)$ as
\begin{equation}\label{eq:dgdt}
	\dfrac{\partial g(V(t),t)}{\partial t}=\frac{g_\infty( V(t))-g(V(t),t)}{\tau},
\end{equation}
and the current $I(t)$ as
\begin{equation}
	I(t)=g(V(t),t)V(t).\label{eq:I}
\end{equation}
Despite the fact that Eq.~(\ref{eq:cond}) needs to be evaluated numerically, we will refer to Eqs.~(\ref{eq:rhos}), (\ref{eq:Iss}) and (\ref{eq:I}) as an analytic approximation (AA) for the voltage-dependent salt concentration profiles, steady-state current, and time-dependent current, respectively. In Sec.\ \ref{sec:FE_ver} we will verify these three equations against full FE calculations of the PNPS equations (\ref{eq:ce})-(\ref{eq:poisson}).

\section{Finite-Element verification}\label{sec:FE_ver}
\begin{figure}[ht]
\centering
     \includegraphics[width=1\textwidth]{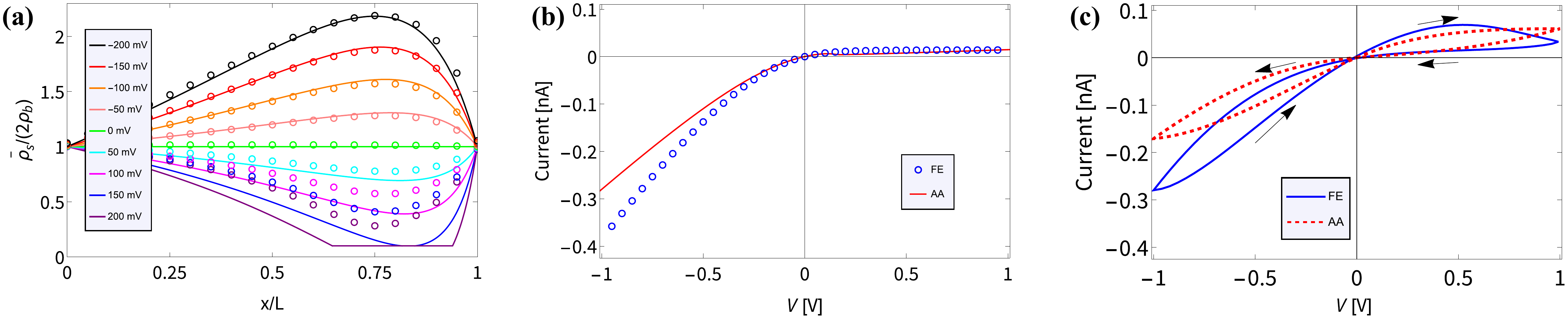}
        \caption{Comparisons of finite-element calculations (FE) of the full PNPS equations (\ref{eq:ce})-(\ref{eq:poisson}) and our analytic approximation (AA) of Eqs.~(\ref{eq:rhos}), (\ref{eq:Iss}) and (\ref{eq:I}), all for our standard parameter set of a bipolar conical channel (see text). \textbf{(a)} The radially averaged salt concentration profiles as determined by Eq.~(\ref{eq:rhos}) (solid lines) and by the FE calculation (circles) for various static potentials $V\in[-200,200]$ mV as indicated by the colours. \textbf{(b)} Steady-state current-voltage relation as predicted by our AA of Eq.~(\ref{eq:Iss}) (red) and by the FE calculations (blue), featuring strong (diodic) current rectification. \textbf{(c)} Current-voltage diagram for an applied periodic triangle potential $V(t)$ with amplitudes $\pm 1 \text{ V}$ and frequency $f=45\text{ Hz}$, revealing a clear pinched hysteresis loop.}
        \label{fig:theorycompare}
\end{figure}

In Sec.\ \ref{sec:AA} we derived an AA for the voltage-dependent salt concentration profiles, steady-state current, and time-dependent current. Here we will verify these results against full FE calculations of the underlying PNPS equations (\ref{eq:ce})-(\ref{eq:poisson}). Throughout this section we will use our standard parameter set and vary the applied voltage. Firstly, in Fig.~\ref{fig:theorycompare}(a) we compare for a variety of positive and negative static voltages $V$ the radially averaged concentration profiles as predicted by Eq.~(\ref{eq:rhos}) (solid lines) with the FE calculations (circles). For $V<0$ we observe ion accumulation and excellent agreement with almost indistinguishable results for AA and FE. For $V>0$ the agreement is still reasonable and qualitative, however a quantitative discrepancy is now clearly visible, especially at larger positive voltages. Whereas the FE concentration profile at the highest voltage ($V=200$ mV, purple circles) shows a depletion of salt down to about 30\% of the bulk concentration at $x/L\simeq 0.75$, the FE-generated concentration at this point remains strictly positive, of course. By contrast, the corresponding AA profile (purple line) falls below 10\% of the bulk concentration (indicated by the horizontal line) and in fact even becomes negative in a neighborhood of $x/L\simeq 0.75$. As we stated before, the extremely low local salt concentration at high $V$ causes a break-down of the AA-assumption of a small Debye length $\lambda_{\ch{D}}$ (compared to $R_t$), a problem that we cure in an ad hoc fashion by replacing $\overline{\rho}_{\ch{s}}$ by $\max\left[0.2\rho_b,\overline{\rho}_{\ch{s}}(x,V)\right]$ in Eq.~(\ref{eq:cond}).

In Fig.~\ref{fig:theorycompare}(b) we translate the concentration profiles at static potentials $V$ to the steady-state current-voltage relation $I(V)$ through Eqs.~(\ref{eq:cond}) and (\ref{eq:Iss}) (red) and compare them with $I(V)$ as obtained from FE calculations (blue). There is a good agreement, most notably a very similar strongly diodic effect is found through both methods, with quantitatively similar currents. The agreement also seems to hold for strong positive potentials, despite the aforementioned decrease of accuracy of the AA for this voltage regime.
\onecolumngrid\
\twocolumngrid\
 We propose that the I-V relation still matches well since the prediction that the channel is locally nearly completely depleted of salt for high static potentials does in fact match with FE calculations. Therefore, replacing $\overline{\rho}_{\ch{s}}(x,V)$ by $\max\left[0.2\rho_b,\overline{\rho}_{\ch{s}}(x,V)\right]$ in Eq.~(\ref{eq:cond}) effectively captures, for this parameter set at least, this depletion and ensures an I-V relation agreement over a wider voltage range than perhaps could have been expected. We do note that the circuit we propose in Sec.~\ref{sec:HHcircuit} relies on potentials in the range $\pm 0.2\text{ V}$, therefore operating on voltages within the AA range of validity.

Lastly, in Fig.~\ref{fig:theorycompare}(c) we plot the current-voltage relation $I(t)$-$V(t)$ for the case of an applied periodic triangle potential $V(t)$ with amplitudes $\pm 1 \text{ V}$ and frequency $f=45\text{ Hz}$, is in line with the prediction that $\tau f\approx 0.19$ yields the most pronounced memory effect \cite{kamsma2023iontronic}. We compare the time-dependent current determined through Eq.~(\ref{eq:I}) (red) against FE calculations (blue). In both instances a similar pinched hysteresis loop is found, the hallmark of a memristor \cite{chua2014if}. We note that this hysteresis loop shows a much more pronounced opening compared to a loop of a similar UP channel \cite{kamsma2023iontronic}, showing that the stronger current rectification of BP channels translates to a stronger memristive effect.

\begin{figure}[ht!]
	\centering
	\includegraphics[width=0.42\textwidth]{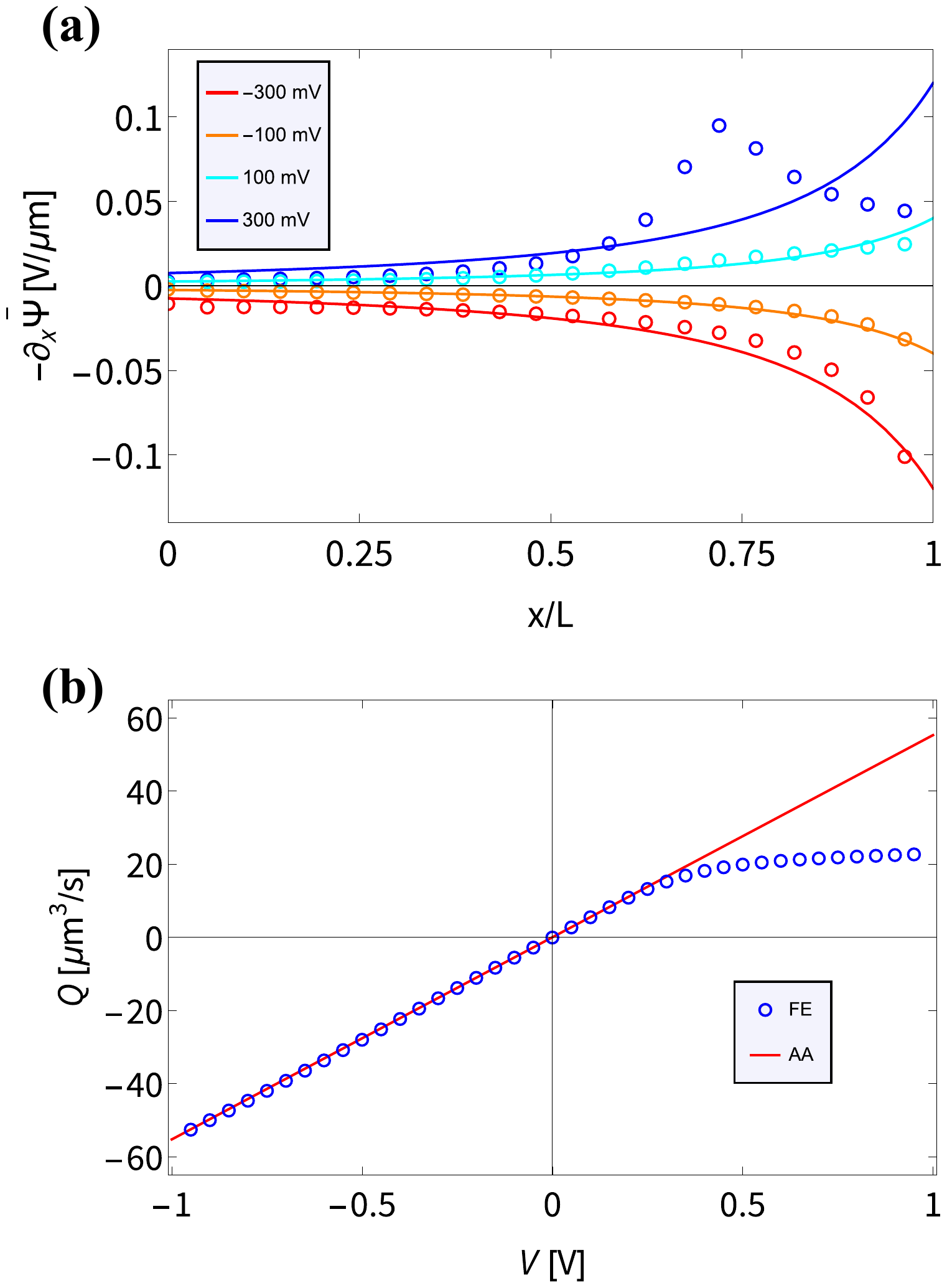}
	\caption{\textbf{(a)} Steady-state electric field $-\partial_x\overline{\Psi}(x)$ inside the channel as predicted by Eq.~(\ref{eq:Efield}) (solid lines) and as measured on the central axis of the channel through the FE calculations (circles) of the PNPS equations (\ref{eq:ce})-(\ref{eq:poisson}) for various static applied potentials $V$. \textbf{(b)} Steady-state fluid volume flow $Q(V)$ as a function of the static potential $V$ as predicted by $Q(V)=-\pi R_{\mathrm{t}}R_{\mathrm{b}}\epsilon\psi_{\ch{eff}}V/(\eta L)$ with $\psi_{\ch{eff}}=-25\text{ mV}$ a fit parameter for the linear regime ($V\lesssim 0.4\text{ V}$) of $Q(V)$ (red) and as determined by FE calculations (blue).}
	\label{fig:EfieldandQ}
\end{figure}

Before we consider iontronic circuits of BP conical channels in Sec.~\ref{sec:HHcircuit}, which essentially only involve the AA approximation of the current-voltage relation, let us consider to what extent the radially averaged electric field $-\partial_x\overline{\psi}(x)$ and the fluid flow $Q(V)$ are accurately described by our AA for BP channels for various static $V$. In the AA $-\partial_x\overline{\psi}(x)$ is given by Eq.~(\ref{eq:Efield}), which shows good agreement for UP conical channels in the present parameter regime \cite{boon2021nonlinear}. In Fig.~\ref{fig:EfieldandQ}(a) we compare the electric field for various static potentials $V$ as predicted by Eq.~(\ref{eq:Efield}) (solid lines) with FE calculations (circles). For negative and moderately positive potentials we find good agreement, as expected on the agreement we found in Fig.~\ref{fig:theorycompare}(a), however a clear disagreement is observed for larger positive static voltages $V\gtrsim 0.2 \text{ V}$. As before, we expect this to be due to the strong ion depletion at high positive potentials, typically in the vicinity of $x\approx2L/3$. The resulting overlapping EDLs in combination with the longitudinal dependence of $\sigma(x)$ create a local buildup of a longitudinally varying ionic charge density, creating a peak in the (no longer divergent-free) electric field around the location of the strongest depletion at $x\approx2L/3$. This explanation relies on the longitudinal electric field within the EDL that is inherently present in BP channels due to the surface charge inhomogeneity; this longitudinal field is not present in UP channels with similar parameters as the surface charge is homogeneous. Moreover, the salt depletion is much weaker in UP channels and thus the underlying assumption that $\lambda_{\ch{D}}\ll R(x)$ remains valid for a wider voltage regime \cite{boon2021nonlinear,kamsma2023iontronic}. This is probably why the peak in $-\partial_x\overline{\psi}(x)$ for $V=300$\,mV in Fig.3(a) is not observed in UP channels in a similar parameter regime \cite{boon2021nonlinear}.

The underlying Eq.~(\ref{eq:rhos}) of the reported result is dependent on the fluid volume flow $Q(V)=-\pi R_{\mathrm{t}}R_{\mathrm{b}}\epsilon\psi_{\ch{eff}}V/(\eta L)$, which we show in Fig.~\ref{fig:EfieldandQ}(b) (red) compared to FE calculations (blue). The relation of fluid flow $Q(V)$ to surface potential $\psi_{\ch{eff}}$ is not immediately clear. In UP channels, $Q(V)\propto\psi_0$ with $\psi_0$ the (homogeneous) surface potential \cite{boon2021nonlinear}, but in BP channels such a relation is not obvious as the surface potential $\psi_0(x)$ is inhomogeneous. In Fig.~\ref{fig:EfieldandQ}(b) we show that using $\psi_{\ch{eff}}=-25\text{ mV}$ as a fit parameter based on the linear regime of $Q(V)$ (i.e.\ for $V\lesssim 0.4\text{ V}$) yields good agreement (red) with FE calculations (blue) for roughly the same voltage regime as where we find good agreement for the electric field. Fascinatingly, from Fig.~\ref{fig:EfieldandQ}(b) we conclude for stronger positive potentials $V\gtrsim 0.4\text{ V}$ that the BP channel acts as a fluidic diode. Remarkably, the tip polarity (here negative) determines the direction of the electro-osmotic flow, positive for positive $V$ and negative for negative $V$, despite the majority of the channel carrying a positive surface charge. Additionally, also the strength of $\psi_{\ch{eff}}=-25\text{ mV}$ seems to be similar to the average surface potential of the tip $\int_{2/3L}^{L}\psi_{0}(x)\dd x/(L/3)\approx -34\text{ mV}$. Whether the tip polarity is a general predictor for the strength and direction of the electro-osmotic flow and whether the fluidic diode behaviour emerges for other parameter configurations requires a more extensive investigation of the parameter space. We leave this topic for future studies and instead here focus on our standard parameter set in order to continue with investigating the iontronic neuromorphic circuit in Sec.~\ref{sec:HHcircuit}.

We conclude this section by stating that although we find deviations for the salt concentration profiles, electric field profiles, and fluid volume flow for relatively large positive potentials, these deviations seem to have a limited impact on the overall $I(V)$ relations as demonstrated in Figs.~\ref{fig:theorycompare}(b) and \ref{fig:theorycompare}(c), which are most relevant in the context of iontronic circuitry. Furthermore, the iontronic circuit presented in Sec.~\ref{sec:HHcircuit} operates within a voltage regime where the electric fields and fluid flows predicted by the analytic approximation are reasonably consistent with FE calculations.

\section{Neuromorphic microfluidic circuit}\label{sec:HHcircuit}
We proceed to investigate the use of BP channels in iontronic circuits, specifically we are interested in neuromorphic circuits. In biological systems the process of neuronal signaling is enabled by the transport of various ionic species through the neuronal cell membrane. Upon a stimulus of sufficient strength and duration a process is set in motion which results in a voltage spike over the membrane due to modulated ionic charge transport through biological ion channels. Such voltage spikes are known as action potentials (APs) and follow the characteristic all-or-none law, meaning that the membrane does not spike at all for stimuli below a critical threshold \cite{lucas1909all,bean2007action,fundNeuroE}. Neurons are also able to generate a series of APs, known as a spike train, which plays a vital role in neuronal communication \cite{fundNeuroTrain,cymbalyuk2002bursting,marder2002cellular,sherman2001tonic,bean2007action}. Inspired by Hodgkin-Huxley (HH) circuits \cite{hodgkin1952quantitative, rall2011core,fitzhugh1973dimensional,rall1962theory,halter1991distributed,hay2011models,hines1997neuron,kole2008action}, developed by treating the neuronal membrane as a circuit \cite{hodgkin1952quantitative}, some iontronic HH circuits were proposed that reproduce neuronal spiking features \cite{kamsma2023iontronic,robin2021principles}, where the circuit in Ref.~\cite{kamsma2023iontronic} applies UP conical channels. Since the BP channels of interest in this manuscript show more pronounced memristive properties compared to UP channels, we expect to be able to improve upon the circuit described in Ref.~\cite{kamsma2023iontronic} by considering parameters that are experimentally more accessible and closer to their biological analogues.

\begin{figure}[ht]
	\centering
	\includegraphics[width=1\textwidth]{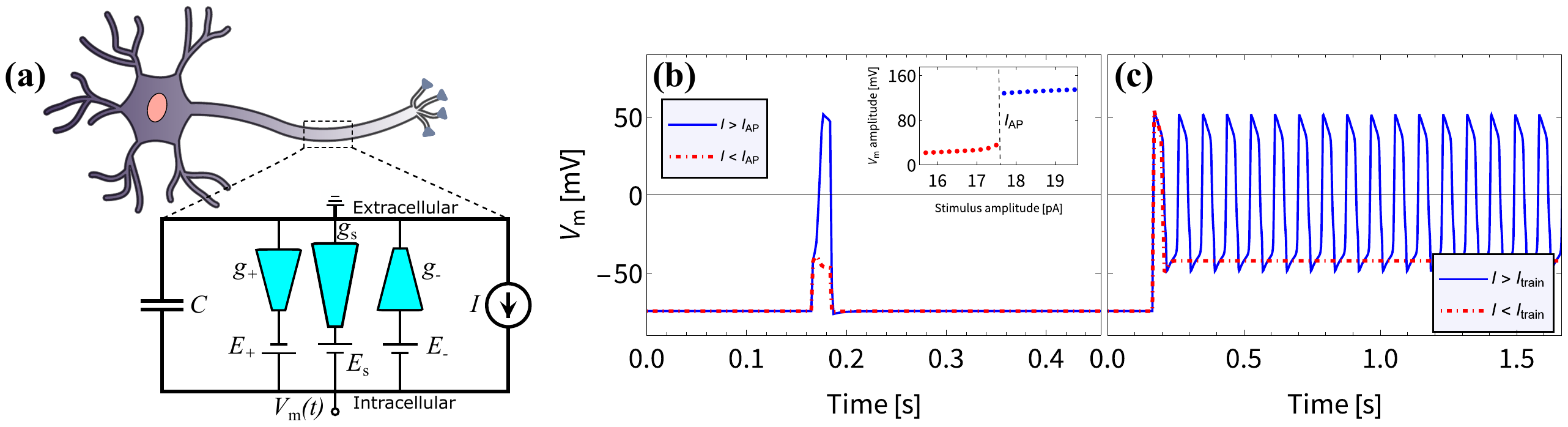}
	\caption{\textbf{(a)} Schematic representation of the circuit proposed in Ref.~\cite{kamsma2023iontronic}, however now with three bipolar rather than three unipolar channels, connected in series to individual batteries and in parallel to a capacitor. The electric potential difference $V_{\ch{m}}(t)$ over the capacitor can be driven by an imposed stimulus current $I(t)$. \textbf{(b)} The membrane potential $V_{\mathrm{m}}(t)$ resulting from an imposed subcritical (red) and supercritical (blue) current pulse $I(t)$ of duration 20 ms and strengths $17.5\text{ pA}$ and $17.6\text{ pA}$ respectively, as determined by Eq.~(\hyref{eq:HH_3_2}), displaying an all-or-none action potential, as can be seen by the jump in spike amplitude around $I_{\mathrm{AP}}=17.5$ pA as shown in the inset. \textbf{(c)} The membrane potential $V_{\mathrm{m}}(t)$ as a result of an imposed subcritical (red) and supercritical (blue) sustained currents $I(t)$  of strengths $18\text{ pA}$ and $18.1\text{ pA}$ respectively, where a spike train emerges for $I(t)>I_{\ch{train}}=18$ pA. The magnitude of the membrane potentials before and during the APs are similar to those observed in mammalian APs \cite{bean2007action}.}
	\label{fig:neurofigs}
\end{figure}

In an attempt to reproduce the all-or-none APs and the spike train found in biological neurons and in the iontronic circuit in Ref.~\cite{kamsma2023iontronic}, we consider the circuit architecture presented in Ref.~\cite{kamsma2023iontronic}, shown in Fig.~\ref{fig:neurofigs}(a), where we replace the UP channels with BP channels with conductances $g_{+}$, $g_{-}$ and $g_{\mathrm{s}}$ and consider a new set of circuit parameters. To separate out the response times of these channels we set the channel lengths to be $L_{\pm}=1\text{ }\mu\text{m}$ and $L_{\mathrm{s}}=15\text{ }\mu\text{m}$. Through Eq.~(\ref{eq:ts}) this translates to $\tau_{\pm}\approx 0.042$ ms for the two fast channels, while the slow channel has a timescale $\tau_{\mathrm{s}}\approx 9.4 \text{ ms } \gg\tau_{\pm}$. The batteries, with which the conical channels are connected in series, have potentials $E_{\pm}=\pm 114\text{ mV}$ for the two fast channels and $E_{\mathrm{s}}=-180\text{ mV}$ for the slow channel. These batteries are the circuit analogues of the Nernst potentials due to concentration gradients over neuronal membranes, where we note that these battery potentials are within the range of typical mammalian Nernst potentials \cite{fundNeuroE}. Moreover, the bulk concentration of $\rho_{\ch{b}}=2\text{ mM}$ is close to typical mammalian extracellular $\text{K}^{+}$ concentrations \cite{fundNeuroE}. A capacitor is connected in parallel to the channels, with a capacitance $C=0.05\text{ }\text{pF}$ that again is close to typical biological values, as this corresponds to the capacitance of mammalian neuronal membrane of area $\sim 2-5\;\mu\text{m}^2$ \cite{gentet2000direct,niles1988planar,solsona1998regulation,sukhorukov1993hypotonically,major1994detailed,thurbon1998passive,chitwood1999passive}, which is of similar dimensions as the surface area of the channels. 

The electric potential $V_{\mathrm{m}}(t)$ over the circuit shown in Fig. \ref{fig:neurofigs}(a) is equivalent to the membrane potential over a neuronal membrane \cite{hodgkin1952quantitative} and responds to the imposed stimulus current $I(t)$, which acts as the control parameter and determines whether spiking occurs. The time-evolution of $V_{\mathrm{m}}(t)$ is provided by Kirchhoff's law
\begin{align}
	C\dfrac{\dd V_{\mathrm{m}}(t)}{\dd t}=I(t)-\sum_{i\in\left\{+,-,\mathrm{s}\right\}}g_{i}(V_i(t),t)\left(V_{\mathrm{m}}(t)-E_{i}\right),\label{eq:HH_3_2}
\end{align}
where the conductances $g_i(V_i(t),t)$ each evolve according to Eq.~(\ref{eq:dgdt}) with the corresponding $\tau_i$. The voltage arguments $V_i$ of $g_{i,\infty}(V_i)$ are given by $V_-(t)=V_{\mathrm{m}}(t)-E_-$, $V_+(t)=-V_{\mathrm{m}}(t)+E_+$ and $V_{\mathrm{s}}(t)=-V_{\mathrm{m}}(t)+E_{\mathrm{s}}$, with the different signs of the potentials corresponding to the different orientations of the channels as depicted in Fig.~\ref{fig:neurofigs}(a). Eqs.~(\ref{eq:ts}), (\ref{eq:dgdt}) and (\ref{eq:HH_3_2}) form a closed set, which we numerically solve with initial conditions $V(0)=-0.1\text{ V}$ and $g_{i}(0)=g_{0,i}$.

\onecolumngrid\
\twocolumngrid\
In Figs.~\ref{fig:neurofigs}(b) and \ref{fig:neurofigs}(c) we show that we reproduce the same neuronal behaviour as found in Ref.~\cite{kamsma2023iontronic}, in the form of all-or-none action potentials (Fig.~\ref{fig:neurofigs}(b)) and a spike train (Fig.~\ref{fig:neurofigs}(c)).  Excitingly, the membrane potentials before and during the APs range from $\sim-70\text{ mV}$ to $\sim50\text{ mV}$ and are therefore of similar magnitude to those observed in mammalian APs \cite{bean2007action}. This, combined with the biologically more relevant battery potentials $E_i$ and bulk concentration $\rho_{\ch{b}}$ compared to the circuit with UP channels from Ref.~\cite{kamsma2023iontronic}, may prove to be crucial for the integration of such an iontronic circuit with biological systems in future applications.

\section{Conclusion and outlook}\label{sec:concl}
In summary, we have presented a theoretical approximation of the voltage-dependent steady-state current and the dynamic conductive properties of conical channels that are filled with an aqueous electrolyte and carry an inhomogeneous surface charge. Specifically, we focus on a channel with a positive surface charge at the base and middle, and a negative surface charge at the tip, thus forming a bipolar channel. This channel exhibits significantly improved current rectification when compared to unipolar conical channels with homogeneous surface charges and otherwise identical parameters. For negative and moderately positive static potentials $V\lesssim 0.2\text{ V}$, our analytic approximation of salt concentration profiles and time-dependent currents are found to be in good agreement with finite-element calculations of the PNPS equations (\ref{eq:ce})-(\ref{eq:poisson}), providing a solid foundation for further investigation of the use of these channels in (neuromorphic) iontronic circuits. While the steady-state and time-dependent current-voltage relations also show good agreement for large potentials, we do observe some qualitative deviations for $V\gtrsim 0.2\text{ V}$ in the salt concentration profiles and electric field profiles compared to finite-element calculations. Additionally, for large static potentials $V\gtrsim 0.4\text{ V}$ we observed a non-linearity in the relation of fluid volume flow and applied potential, where the bipolar channel acts as a fluidic diode. We hypothesize that this is due to the strong salt depletion that bipolar channels exhibit at large potentials, which implies that the small-Debye-length assumption $\lambda_{\ch{D}}\ll R(x)$ that underlies Eq.~(\ref{eq:rhos}) becomes increasingly less accurate. Although the microscopic salt concentration profiles and electric field profiles are not accurately predicted by the present analytical model, the overall steady-state and time-dependent conductance is still in good agreement, indicating that our presented analytical approximation is an effective tool for the exploration of bipolar channels for iontronic circuits.

By extending the analytical methodology of Refs.~\cite{kamsma2023iontronic,boon2021nonlinear} to bipolar conical channels, we have demonstrated its generalizability and potential for predicting features such as current rectification in a wider range of geometries and surface charge distributions. Our derived equations suggest that the model we present here may be directly applicable to predicting current rectification in bipolar cylindrical channels, rather than solely conical geometries, which is previously experimentally demonstrated \cite{karnik2007rectification,daiguji2005nanofluidic,meng2015cooperative}. Furthermore, since our model allows for any any general linear increase in surface charge along the longitudinal axis, our approach may also aid in identifying optimized surface charge values, distributions, and geometries for iontronic systems, beyond the parameter set on which we focus in this work. These findings point towards the generality, utility and potential of this analytical methodology in the field of iontronics.

In addition to the implications for optimizing and understanding individual channel properties, this work has also highlighted the potential of this analytic approximation method in the context of exploring iontronic circuits. By modeling a Hodgkin-Huxley circuit with bipolar channels we are able to present a system that relies on battery potentials and on salt concentrations comparable to their biological analogues, and which produces all-or-none action potentials and spike trains with voltage membranes that closely resemble the values observed in biological systems. This suggests that further research in this direction may prove beneficial in the development of advanced iontronic devices with improved performance.

\begin{acknowledgments}
This work is part of the D-ITP consortium, a program of the Netherlands Organisation for Scientific Research (NWO) that is funded by the Dutch Ministry of Education, Culture and Science (OCW). T.M.K.\ performed the calculations; T.M.K.\ and W.Q.B.\ conceptualized the work; T.M.K. and W.Q.B. developed the theory under supervision of C.S.\ and R.v.R..
\end{acknowledgments}

\bibliographystyle{rsc}

%

\end{document}